\documentclass[journal=ancac3,manuscript=article]{achemso}

\usepackage[version=3]{mhchem} 
\usepackage{siunitx}

\author{Guillaume Graciani}
\affiliation{Institute for Basic Science - Center for Soft and Living Matter, Ulsan, South Korea}
\author{John T. King}
\affiliation{Institute for Basic Science - Center for Soft and Living Matter, Ulsan, South Korea}
\author{Fran\c{c}ois Amblard}
\affiliation{Department of Physics, Ulsan National Institute of Science and Technology, Ulsan, South Korea}
\email{famblard@protonmail.com}

\title[An \textsf{achemso} demo]
  {Cavity Amplified Scattering Spectroscopy reveals the dynamics of
proteins and nanoparticles in quasi-transparent and miniature samples}


\begin{document}






\begin{abstract}
Dynamic light scattering techniques are routinely used for numerous industrial and research applications, because they can give access to the motion spectrum of micro- and nano-objects, and therefore to particle sizes or visco-elastic properties.
However, measurements are impossible when samples do not scatterer light enough, \textit{i.e.} when there are too few scattering events due to excessively small scattering cross-sections and/or low concentrations of scatterers.
Here, we propose to amplify light scattering efficiency by placing weakly scattering samples inside a Lambertian cavity with high reflectance walls. 
It produces a 3D isotropic and homogeneous light field that effectively elongates the scattering pathlength by \numrange{2}{3} orders of magnitude, and leads to a dramatic increase in sensitivity.
We could indeed measure the diffusion coefficient and size of particles ranging from $\SI{5}{\nano\metre}$ to $\SI{20}{\micro\metre}$ with volume fractions as low at $10^{-9}$ in volumes as low as $\SI{100}{\micro\liter}$, and in solvents with refractive index mismatches down to $\Delta n \approx 0.01$.
With a  $10^{4}$ fold increase in sensitivity compared to classical techniques, we considerably expand the applications of light scattering to highly diluted samples, miniaturized microfluidics samples, and samples practically deemed non-scattering.
Beyond the realm of current applications of light scattering techniques, our Cavity Amplified Scattering Spectroscopy method (CASS) and its outstanding sensitivity represent a major methodological step towards the study of problems such as the ballistic limit of Brownian motion, the internal dynamics of proteins, or the low frequency dielectric dynamics of liquids.
\end{abstract}

\section{Introduction}
Dynamic Light Scattering \cite{berne2000dynamic} (DLS) is a powerful technique to probe the dynamics of materials in physics, chemistry and biology, and has countless applications.
Relying on the fundamental work of Stokes \cite{stokes8g} and Einstein \cite{einstein1905erzeugung, einstein1906theorie, einstein1910theory}, it quickly became one of the most popular methods to measure the spectrum of thermal motions of colloidal particles and infer their size distribution, under the provision that samples do not scatter individual photons more than once on average.
Beyond this limit of DLS, Diffusing Wave Spectroscopy (DWS) \cite{maret1987, pine1988diffusing, mackintosh1989diffusing} extends the power of light scattering to highly scattering and possibly non-ergodic media \cite{scheffold2001diffusing}, in which \r{a}ngstrom motions can be detected, under the provision that each photon undergoes a larger number of scattering events when traveling through the sample.
Beyond particle sizing applications, DWS therefore became the main tool to optically measure the rheological properties of gels, foams and emulsions \cite{zakharov2009advances,kim2019diffusing} of interest to various industries, as well as for basic research.
Recently, DWS has also been used to measure the non-thermal fluctuation spectrum of acto-myosin due to the mechano-enzymatic activity of motor proteins \cite{le2001motor,graciani2020protein}.
However, for both DLS and DWS, a motion spectrum cannot be assessed when the intensity of the scattered light is too weak, \textit{i.e.} fundamentally when the sample looks transparent because its size is too small compared to the scattering length of the material. 
If larger samples cannot be prepared, light scattering is simply not an option.

In this paper, we present a method called Cavity Amplified Scattering Spectroscopy (CASS) that fundamentally elongates the effective optical length of samples by a factor $10^2$ to $10^3$, and considerably increases the sensitivity limit of DLS and DWS, allowing for up to $10^4$ folds smaller volume fractions to be probed.
The amplification is obtained by embedding samples inside a closed high-albedo Lambertian cavity \cite{ulbricht1905} injected with a highly coherent laser light.
The walls of the cavity reflect the light and force it back through the sample up to thousands of times, leading to a very long average photon path, that remains shorter than the coherence length of the laser by design.
The light being reflected by Lambertian scattering from the walls, the whole sample volume is immersed into - and probed by - a homogeneous,  isotopic, and coherent optical field.
Because light scattering by the walls is static, the resulting 3D interference pattern (speckle field) will only fluctuate if the scattering sample is dynamic.
Because of the statistical homogeneity of the field, speckle fluctuations provide the same information regardless of where they are measured.
Our CASS setup is particularly efficient for dilute samples, with concentrations orders of magnitude lower than the relatively high concentrations required for DLS measurements, \textit{e.g.} 0.1 to 1 mg/ml for proteins, and volume fractions higher than $10^{-5}$ for micron-sized particles.
The study of colloidal objects with low scattering efficiencies is also possible in CASS.
After a presentation of the CASS method, we compare its sensitivity with classical DLS and DWS, and we show that it allows for the size measurement of particles in highly diluted suspensions or with small refractive index mismatches with their solvent.
Finally, we describe a miniaturized CASS setup that significantly reduces the probe volume, and potentially leads to new research, industrial, and environmental applications of light scattering beyond its current usage.

\section{Results}
\subsection{\label{sec:level3} Concept of 3D stochastic interferometry } 

In a recent report, 3D stochastic interferometry has been introduced as a highly sensitive method to homogeneously probe the geometric and dielectric fluctuations of an optical volume using a high reflectivity diffuse Lambertian cavity \cite{graciani20213d,graciani2019random}.

This method is based on the propagation of a single-frequency laser light inside a closed cavity with diffuse reflective walls, which leads to the creation of a 3D coherent random light field that fills the cavity volume.
Based on the wave particle duality of light, one can consider the cavity to be filled by a homogeneous gas of monochromatic photons individually associated with an isotropically distributed direction.
As a result, at each point inside the cavity, the light corresponds to the coherent superposition of a very large number of interfering waves, leading to a 3D speckle pattern which is stationary unless the geometry of the cavity or the dielectric properties of its content fluctuate. 
The response to such dielectric or geometric perturbations is a change of the intensity of the speckle field with time.
Due to the statistical homogeneity and isotropy of the photon gas, all points in the cavity, all directions and all polarization states are statistically equivalent regardless of the geometry of the cavity \cite{graciani20213d}.
It follows a unique property: statistically, for any two points A and B inside the cavity, the phase perturbation occurring at point A is transported by the random field to point B, and the statistical properties of the random path from A to B do not depend on their positions. 
As a consequence, the cavity and the field inside can be used as a unique interferometer, in the sense that the response to aforementioned geometric or dielectric perturbations depends neither on the cavity shape, nor on how the light is injected, nor on where perturbations take place, nor on where the response is measured \cite{graciani20213d}.

\subsection{\label{sec:level3} Cavity design }

Practically, a highly reflective cavity was used with inside walls made of compressed quartz powder \cite{cone2015diffuse}, or Teflon.
The cavity albedo assessed from the spreading of a light pulse was $\rho=1-\epsilon= 99.94 \pm 0.005\%$, where the loss coefficient $\epsilon$ results from absorption by the walls.
A smaller albedo was found for teflon cavities ($\rho=$99.1\%).
A $\SI{660}{\nano\meter}$ single frequency (linewidth $<$1MHz) 300 mW continuous wave laser was injected inside the cavity through a multimode fiber.
The scattered light was collected by a single-mode fiber split into two single photon counting modules connected to a digital correlator (Fig.\ref{fig:1} (a)). It measures a cross-correlation, which gives the intensity auto-correlation without afterpulsing artifacts. 
Between two successive reflections, each photon travels along a chord, with an average chord length $L_{c}=\frac{4 V_{c}}{\Sigma_{c}}$ (where $V_{c}$ is the cavity volume and $\Sigma_{c}$ is its internal surface  area) \cite{fry2006integrating}, which represents the characteristic linear size of the cavity.
The typical number of reflections before extinction is expressed as 
$n=\frac{1}{-ln(\rho)}$ \cite{fry2016integratingicrds}, which we identify as the amplification gain $g_{0}=n\approx 1/\epsilon$.
Consequently, the typical distance traveled by photons inside the cavity is $L_{0}=L_{c}*g_{0}$ which, with a typical gain measured at $g_{0}=1700$ for quartz cavities, amounts to $\approx \SI{62}{\meter}$, below the $\SI{95}{\meter}$ coherence length of the laser.
If a sample is introduced into the cavity, such that each photon crossing that sample volume has a probability $p_{c}$ ($\ll1$) to be scattered once (DLS) or more times (DWS), the probability to be scattered at least once after traveling $g_{0}$ chords reads $1-(1-p_{c})^{g_{0}}\approx g_{0}p_{c}$.

\subsection{\label{sec:level3} Principle of coherent amplification }

\begin{figure*}
\includegraphics[width=\linewidth]{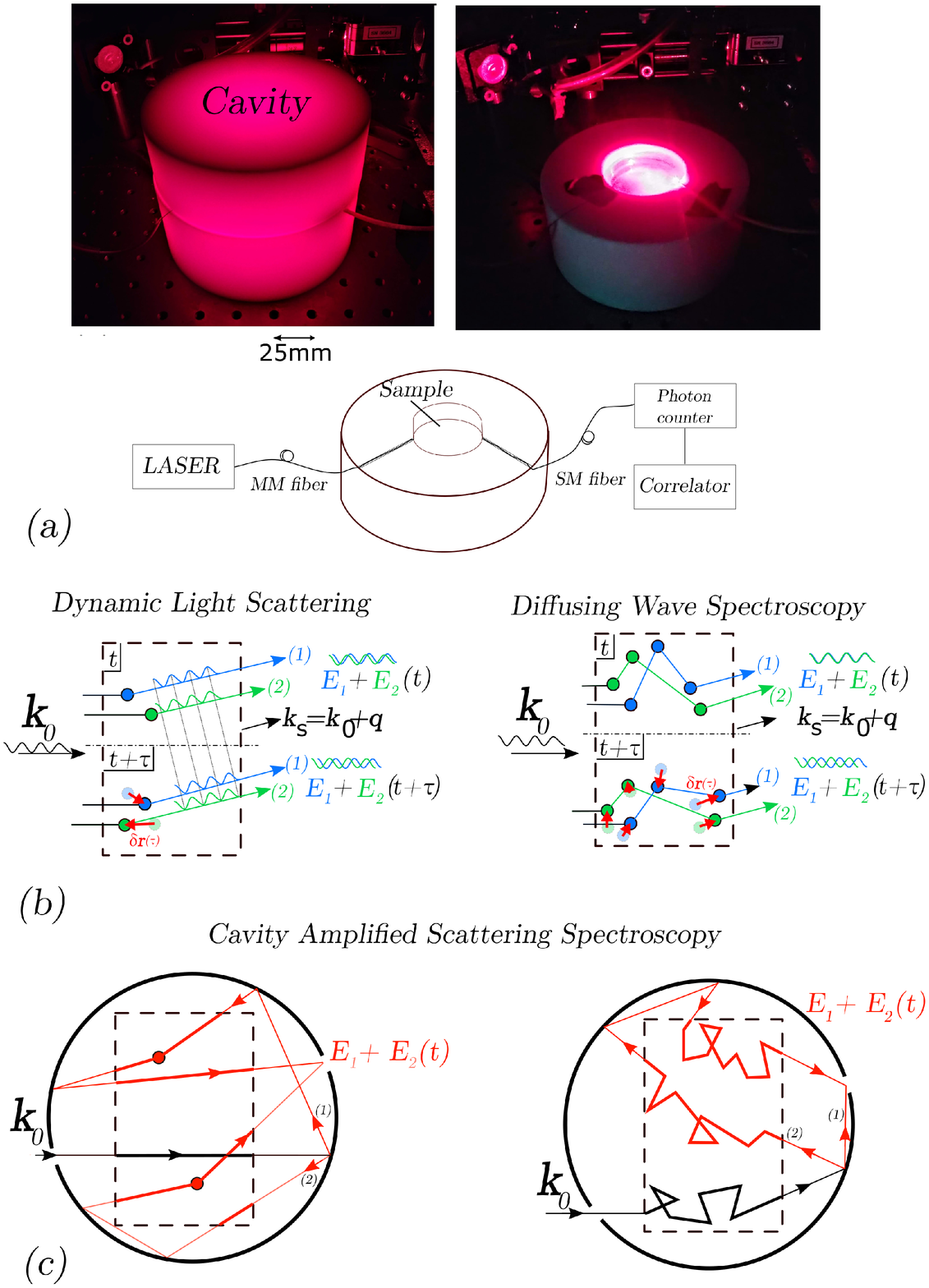}
\caption{\label{fig:1}}
\end{figure*}

The Lambertian cavity described above is in  principle identical to the so-called Ulbricht integrating sphere \cite{ulbricht1905} used in the past century for classical anidolic applications in optics (radiometry, ring-down absorption), and recently for stochastic 3D/2D light concentration \cite{khaoua2021stochastic}.
However, to the best of our knowledge, Ulbricht cavities have only been used for incoherent applications, and our purpose is to use them instead as coherent amplifiers of scattering signals produced by DLS or DWS.
In the DLS method (Fig.\ref{fig:1} (b)), a single input plane wave is considered with wave vector $\vec{k_{0}}$, and the phase of the field scattered by a given scatterer in the direction of the wave vector $\vec{k_{s}}$ changes during the time interval $\tau$ as a function of the random motion $\delta\vec{r}(\tau)$ of that scatterer projected onto the direction perpendicular to the scattering angle vector  $\vec{q}=\vec{k_{s}}-\vec{k_{0}}$. 
The resulting phase variation $\delta\varphi = \vec{q}\cdot \delta \vec{r}$ leads to a variance $\sigma^2_{\delta\varphi(\tau)}$, which drives the decorrelation of the light intensity in the direction $\vec{k_{s}}$ due to the independent superposition of the fields produced by a large number of scattering events.
In DWS, because each photon harbors multiple scattering events, it is considered to follow a Brownian path over which individual phase variations $\delta\varphi_{i} = \vec{q_{i}} \cdot \delta \vec{r_{i}}$  add-up.
The resulting variance $\sigma^2_{\delta\varphi(\tau)}$ of the output phase $\delta\varphi = \Sigma \delta\varphi_{i}$ carried by each Brownian path can be shown to be proportional to the mean square displacement $\sigma^2_{\delta r^{2}(\tau)}$, and to an effective number of scattering events that depends on the path length and the scattering anisotropy.
For both DLS and DWS, the intensity decorrelation at the time-scale $\tau$ is proportional to the elementary position variance $\sigma^2_{\delta r^{2}(\tau)}$ and to the effective number of scattering events that individually and independently contribute to the variance of the phase $\sigma^2_{\delta\varphi(\tau)}$ of the superimposed fields, regardless of the angular dependence considered in DLS.
In short, our cavity simply multiplies this effective number of scattering events by the path amplification gain $g_0$ (Fig.\ref{fig:1} (c)).

\subsection{DLS and DWS amplification}

To demonstrate the concept of CASS and assess how much scattering dynamics is amplified, we compared the normalized auto-correlation functions of the scattered light intensity $|g^{(1)}(\tau)|^{2}$ obtained for both DLS and DWS with their cavity-amplified counterparts, using a teflon cavity with $g_{0}\approx 100$.
DLS being most frequently used for diluted suspensions of scatterers, we first measured the dynamics of \SI{5.5}{\micro\meter} diameter PMMA particles in the single scattering regime (Fig.\ref{fig:2} (a)), \textit{i.e.} with a sample size $l_{sa}$ much smaller than the mean transport length $l^{\star}$ ($l^{\star}/l_{sa}\approx 9$), where $l^{\star}$ is the typical length over which the direction of light becomes statistically isotropic.
As expected for a mono-disperse suspension, the DLS auto-correlation decays exponentially to zero with a single rate, due to the Brownian dynamics $\left< \delta r^{2}(\tau) \right> = 6D \tau$ with diffusion coefficient $D$.
As explained above, the amplification of the phase variance by the cavity leads to a faster intensity decorrelation as if the particles had a larger diffusion coefficient.
As a consequence, the time-scale $\tau_{1/2}$ of half decorrelation shifts from $25$ to  \SI{0.38}{\milli\second}, with a ratio $\approx 76$ that approximately matches $g_{0}$.
At the other end of the turbidity scale, we investigated an SDS-stabilized mono-disperse oil/water emulsion of densely packed droplets (Fig.\ref{fig:2} (b)) \cite{kim2019diffusing}.
This sample produces multiple scattering because its linear size ($l_{sa}\approx \SI{10}{\milli\meter}$) is much larger than the mean transport length $l^{\star}\approx \SI{0.25}{\milli\meter}$ ($l_{sa}/l^{\star}\approx40$).
Dense packing leads to a well characterized visco-elastic behavior, such that the thermally excited motions of droplets $\left< \delta r^{2}(\tau) \right>$ increases linearly at short times ($\tau < \SI{10}{\micro\second}$), and gradually bends and saturates at longer limes ($\tau > \SI{10}{\milli\second}$) with a maximal displacement amplitude of $\SI{900}{\pico\meter}$ \cite{kim2019diffusing}. 
Because the intensity decorrelation caused by motions is mathematically more complex for DWS than DLS, no single exponential is recovered.
Still, our DWS observations are congruent with the known dynamics, with a monotonic decay followed by a plateau with a $1.3\%$ decorrelation amplitude for $\tau \geq \SI{1}{\milli\second}$.
As expected, CASS leads to a larger decorrelation ($6\%$) for the plateau.
At the shortest sampling time-scale $\tau_{s}=\SI{12.5}{\nano\second}$, because the experimental noise of the auto-correlation function is small enough for CASS ($\approx 2.10^{-4}$), the decorrelation can be detected from the smallest observable time-scales $\tau \geq \tau_{s}$.
Since $\sqrt{\left< \delta r^{2}(\tau_{s}) \right>} \approx \SI{25}{\pico\meter}$ \cite{kim2019diffusing}, this typical motion amplitude can be considered as our detection limit in the present context.
We should also note that the auto-correlation noise is significantly larger at short time-scales for DWS ($\approx 10^{-3}$) 
relative to CASS. 
This is explained by the fact that shot noise is dominant at this time scale, since photons are typically counted at a rate $R$, which translates into a Bernouilli counting process with probability $R\tau_s\ll1$ at the sampling time $\tau_s$.
In the cavity, the counting rate ($R\approx \SI{5e5}{\per\second}$) is simply multiplied by the factor $g_0$ and the noise on the intensity is thus reduced by the factor $\sqrt{g_0}$.

\subsection{The parameters of amplification}

Schematically, cavity amplification benefits both to the static and the dynamic properties of light scattering. Namely, it increases the static scattering efficiency of the sample, and it amplifies the fluctuations of the scattered intensity.
The former mechanism is interesting for samples that are too transparent for DLS, while the latter is relevant for both DLS and DWS samples.
The benefits for DWS samples are discussed in the previous section and shown on Fig.\ref{fig:2} (b), but are not the focus of the present work.
For DLS or quasi-transparent samples instead, we want to elucidate quantitatively which parameters contribute to the amplification mechanism.
To this end, we first examined how the albedo determines the intensity auto-correlation produced by a DLS sample, namely a colloidal suspension of \SI{27}{\nano\meter} diameter polystyrene particles.
A series of 5 albedo values ($\rho = 0.978$ to $0.991$) is obtained by tuning the reflection loss ratio 
$\epsilon = 1-\rho$ using known absorbers on the cavity walls.
The albedo was assessed by measuring the time-stretching of a picosecond laser pulse, as typically done with cavity ring-down absorption methods \cite{fry1992}, which provides us with the photon transit half-time $\tau_{tr}$.
The corresponding decorrelation amplitudes $\mathcal{D}(\tau)=1-|g^{(1)}(\tau)|^{2}$ (Fig.\ref{fig:3} (a)) increase with $\rho$ with the scaling factor $g_0\approx 1/\epsilon$, leading to the albedo-independent decorrelation response $\mathcal{D}(\tau)/g_0$.
Moreover, as expected from the homogeneous nature of the coherent photon gas inside the cavity due to the Lambertian reflectivity \cite{graciani2019random}, the localization of the scatterers inside the sample have no effect on their contribution to the intensity auto-correlation, as shown with samples that contain the same total number of particles in different volumes (Fig.\ref{fig:3} (b)). 
Conversely, regardless of the volume, an increase in the total number of scatterers $N_p$ leads to a proportionally faster decorrelation, and the decorrelation signal can be directly normalized by $N_p$ (Fig.\ref{fig:3} (b)), leading to the the $N_p$-independent decorrelation response $\mathcal{D}(\tau)/N_p$.

\subsection{\label{sec:level3} Simple theoretical model for weakly scattering samples}

The main purpose of DLS being particle sizing, the following section explains how the size of colloidal particles can be measured by CASS, using the classical DLS model and the statistics of the amplification process.
The key information in DLS is carried by the phase variation 
$\delta\varphi^2 (\tau,\vec{q}) = \vec{q}\cdot \delta \vec{r}$, which entails a well-know dependence on the scattering angle $\theta$ (Fig.\ref{fig:1} (b)), that is described by the anisotropy factor $g=\left<\cos \theta\right>$.
As a consequence, the variance of the phase $\varphi_{(1)}$ produced by single scattering events with typical mean-square movements $\left<\delta r^2(\tau)\right>$ reads  
$\left<\delta\varphi_{(1)}^{2}(\tau)\right>=2 k_{0}^2 [1-g] \left<\delta r^2(\tau)\right> $, and vanishes when the anisotropy is strong $g\approx1$, \textit{i.e.} $q \approx 0$.
When diluted samples are probed by DLS, the detection limit is quickly reached due to an insufficient intensity being scattered, either because of a strong scattering anisotropy that carries vanishingly little information ($q\approx0$), or because isotropic scattering is produced by too few scatterers.
For high enough dilutions, most of the light is not scattered, regardless of the anisotropy factor. 
While the proportion of the light that is scattered over the path length $l_{sa}$ is controlled by the ratio $p=\l_{sa}/l_{sc} \ll 1$ to the scattering length $l_{sc}$, the contribution of an individual scattering event to the phase variance depends on the anisotropy and is weighted by $[1-g]$, as if the scattering efficiency ratio $p$ should be substituted by an effective scattering ratio $p^\star=p[1-g]$.
It comes that $p^\star=l_{sa}/l^{\star}$ where $l^{\star}=l_{sc}/[1-g]$ is the so-called mean transport length, \textit{i.e.} the effective scattering length corrected for by the anisotropy.
What matters indeed is the length $l^{\star}$ the light needs to travel before its direction is randomized compared to the input direction.

Back to the particle sizing problem, which basically amounts to inferring the hydrodynamic radius from the diffusion coefficient $D$ that drives the Brownian dynamics $\left<\delta r^2(\tau)\right> =6D\tau$, the diffusion coefficient is classically obtained by fitting the DLS intensity normalized auto-correlation curve $G_{dls}(\theta,\tau)$ by the equation: 
\begin{equation}{\label{exponentiel beta G_DLS}}
    G_{dls}(\theta,\tau) 
    = \exp [-\beta_{dls}(\theta)\tau]
\end{equation} 
with $\beta_{dls}(\theta)=2q^2(\theta)D$.
For CASS, the auto-correlation decays as $\exp[-\left<\delta\varphi_{(tot)}^{2}(\tau)\right>/2]$, where the variance of the total phase reads 
$\left<\delta\varphi_{(tot)}^{2}(\tau)\right>=g_e p \left<\delta\varphi_{(1)}^{2}(\tau)\right>$, with an effective cavity gain  $g_e$ obtained by correcting the bare gain $g_0$ by the effect of sample absorption (see Supplementary Materials).
As a result, one simply needs to fit the experimental auto-correlation curve with:
\begin{equation}{\label{exponentiel beta CASS}}
    G_{cass}(\tau) 
    = \exp [-\beta_{cass} \tau]
\end{equation} 
with $\beta_{cass} = 6 g_{e} p^{\star} k_{0}^{2} D$.
The relevant mathematical tools are identical for DLS and CASS when dealing with fitting the multi-exponential decays observed for non-monodisperse suspensions (conventional cumulant or CONTIN \cite{ju1992contin} analysis).
This simple model is in agreement with the experimental data shown on Figure \ref{fig:3} (d). Indeed, the CASS auto-correlation signal measured for a suspension of \SI{20}{\micro\meter} diameter PMMA particles (black circles) overlaps rather well with the theoretical DLS auto-correlation (blue line) computed from equation \ref{exponentiel beta G_DLS} for $\theta=\pi/2$, and re-normalized by the CASS enhancement factor (red line):
\begin{equation}{\label{CASS enhancement}}
    \beta_{cass}/\beta_{dls}^{(\theta=\pi/2)} =3 g_e l_{sa}/2l^{\star}
\end{equation}

\subsection{\label{sec:level3} Performance of CASS beyond the sensitivity limit of DLS}

To demonstrate how far CASS can operate beyond the sensitivity limit of DLS, we report now on the performance of particle sizing for quasi-transparent samples, \textit{i.e.} samples that contain too few scatterers or whose scattering cross-section $\sigma_{sc}$ is too small.
A series of measurements were made with CASS and with a commercial DLS particle sizer, for particles ranging from $\SI{5}{\nano\meter}$ to $\SI{20}{\micro\meter}$ in diameter (Fig.\ref{fig:4}(b)).
As expected, we find that CASS operates with suspensions that are up to $10^4$ times more diluted that the dilution limit of commercial DLS. This sensitivity limit is reached in classical DLS when the signal-to-noise ratio is so small that the auto-correlation comes with an excessive noise, and the fitting procedure yields an excessive error on the particle size.
Experimentally for DLS, $\textrm{SNR}\geq 100$ leads to a \numrange{2}{4}$\%$ error, while the error exceeds $10\%$ when $\textrm{SNR}\leq 10$.
Our observation raises two major questions that we address now, namely what determines the SNR, and why the DLS sensitivity and the benefit of CASS depend on the specific properties of each suspension, namely particle size and refractive index.

The SNR for both DLS and CASS first depends on the amplitude of the signal produced by phase fluctuations, which only depend on the mean transport length $l^\star$ and the diffusion coefficient $D$, provided the effective gain $g_{e}$ is unchanged.
Meanwhile, the SNR suffers in both cases from the same sources of noise (intensity noise of the laser, stray light, internal reflection, detector noises, digital correlation noises \cite{graciani20213d}).
In addition, due to its enhanced sensitivity, CASS is also affected by other noise sources such as seismic and acoustic noise, temperature fluctuations, or laser frequency noise \cite{graciani20213d}.
DLS noise is measured as the time-averaged intensity scattered at a right angle by an empty cuvette (or pure water sample for scattering particle suspensions), and typically amounts $\approx 2.10^3 \textrm{photons/s}$ for our high-end instrument.
CASS noise instead depends on $\tau$ and is defined as the standard deviation $\sigma_{|g^{(1)}|^2}(\tau)$ of $G_{cass}(\tau)$. 
Its typical value in our sizing experiments is $\approx 10^{-4}$ (black line on Fig.\ref{fig:4} (c)) and it appears as the noise floor spectrum for the curves shown on Fig.\ref{fig:4} (a).
Obviously, DLS and CASS noise depend on the measurement times, which were identical in both experimental conditions.
The dramatic improvement of the SNR leads to a $10^2$ fold increase of the temporal dynamic range, as illustrated on figure (Fig.\ref{fig:4} (a)).
For the same volume fraction ($10^{-5}$), the decorrelation signal given by CASS exceeds the noise for $\tau\geq\SI{1}{\micro\second}$ while this only happens for $\tau\geq\SI{200}{\micro\second}$ in DLS.
For a dilution $10^{2}$ times beyond the DLS limit, the CASS decorrelation signal exceeds the noise from $\tau \geq \SI{20}{\micro\second}$, and it safely gives the particle size with a $2\%$ uncertainty.
Even further beyond, with a $10^{4}$ times larger dilution, the size is obtained with a $8\%$ uncertainty. 

Why the sensitivity of DLS changes with the size and the nature of the particles is only accounted for by the interplay of $l^{\star}$ and $D$, since the DLS noise is size-independent by definition.
Why the sensitivity enhancement observed with CASS depends on the nature and size of the particles is explained both by the structure of the enhancement factor $\beta_{cass}/\beta_{dls}^{(\theta=\pi/2)}$ given by equation \eqref{CASS enhancement} and by the $\tau$-dependence of the noise $\sigma_{|g^{(1)}|^2}(\tau)$ (Fig.\ref{fig:4} (c)).
Indeed, for smaller particles, the fit is made at shorter time-scales which correspond to a larger noise. Moreover, we observe a reduced sensitivity enhancement for gold particles due to a smaller absorption length of the sample (Fig.\ref{fig:4} (b)).
Finally, DLS does not work for typical protein solutions, unless one uses high enough concentrations and/or large samples. The lowest recommended concentration for most commercial instruments (which brings an error of about $7\%$ on the size estimation) is $\SI{0.1}{\milli\gram\per\milli\liter}$ in a typical cuvette, simply because the scattering cross-section of proteins is too small.
These limitations are overcome by CASS, as shown with figure \ref{fig:4}, with the sizing of the globular protein lysozyme (15 kDa and \SI{1.9}{\nano\meter} radius).
Beyond the DLS sensitivity limit, CASS measurement comes with a $6\%$ error for a 10-fold lower concentration, and $16\%$ error for a 100-fold lower concentration (\SI{10}{\micro\gram\per\milli\liter}).

We should here point out that most commercial DLS instruments are designed to increase the signal by focusing the input laser on probe volumes in the order nanoliters, with linear sizes in the order of a few tens of microns in the direction perpendicular to the beam.
As a consequence of particles randomly entering and exiting such small probe volumes, the number of scatterers fluctuates, and these fluctuations incoherently contribute to the intensity decorrelation as if the laser intensity was unstable. 
The second consequence of reducing the probe volume, is that only pairs of close-by scatterers contained in the probe volume will contribute to the scattering dynamics.
This exclusive selection of short distance pairs introduces a bias if the motion of scatterers are correlated through distance-dependent interactions, such as hydrodynamic interactions.
There is no such short-distance bias or number fluctuations in CASS because it probes the whole sample volume.

Finally, if a pair of scatterers undergoes strongly correlated motions ($\delta \vec{r_{1}}=\delta \vec{r_{2}}$) it will not contribute to the variation of the scattered intensity in DLS, but it will in CASS.
Indeed, since the CASS signal is built upon averaging on all scattering vectors $\vec{q}$, the motions $\delta \vec{r_{1}}$ and $\delta \vec{r_{2}}$ will independently contribute to the dynamics regardless of their correlation.

\subsection{Miniaturization}\label{sec:mini} 

The benefit of CASS makes it possible to perform dynamic scattering measurements on miniature samples in microfluidic-like conditions.
Indeed, the CASS enhancement factor (equation \ref{CASS enhancement}) means that the gain translates either into an effective mean transport length $l^{\star} \mapsto [l^{\star}/g_{e}]$ smaller than the actual one $l^{\star}$, 
or an effective sample size $ l_{sa} \mapsto [g_{e}l_{sa}]$ larger than $l_{sa}$.
A miniature setup was built ($\mu$-CASS) using a 100$\mu$L teflon cavity with a 99.1\% albedo (Fig.\ref{fig:5} (a)). 
This microcavity was filled with a 1$\%$ volume fraction suspension of \SI{27}{\nano\meter} diameter Polystyrene particles, and the CASS decorrelation signal was compared to the 90$^\circ$ angle DLS signal obtained with a homemade setup, using a standard $l_{sa}=\SI{1}{\milli\meter}$ path cuvette filled with the same $\SI{100}{\micro\liter}$ sample.
Our measurements show that the DLS signal was barely above the detector noise level (2000 \textit{v.s.} 1500 counts/s), while CASS yielded a 100-fold larger decorrelation signal ($0.2$ \textit{v.s.} $0.003$ for DLS) at comparable time-scales (Fig.\ref{fig:5} (b) and (c)). 
Teflon was used here for commodity, but an order of magnitude larger decorrelation is expected with high albedo (99.94$\%$) fumed Silica powder cavity instead \cite{graciani20213d}. 
The major gain in sensitivity we observed arises from the very fact that the whole sample is probed by CASS, while DLS probed a $10^{5}$ smaller volume. 
This is indeed typical of classical DLS instruments as explained in the previous section.
In addition, because of the homogeneous sensing provided by CASS, the microfluidic cavity can be designed with little constraints on its shape and the geometry of light injection and detection.

\subsection{\label{sec:level0}Small index mismatch}

The theory of light scattering tells us that the scattering efficiency vanishes when the mismatch $\Delta n$ between the refractive indices of the scatterers and their surrounding medium is too small.
To demonstrate how CASS performs in small mismatch conditions, we used suspensions of \SI{5}{\micro\meter} PMMA particles ($n_{p}\approx1.488$) in water ($n_{w}\approx 1.333$) and oil ($n_{o}\approx 1.474$), and with a small series of volume fractions $\phi = 10^{-1}$, $10^{-2}$ and $10^{-3}$.
From Mie scattering theory, reducing the mismatch from water ($\Delta n_{w}\approx 0.155$) to oil ($\Delta n_{o}\approx 0.014$) increases the mean transport length $l^{\star}$ by the factor 192.
Using our commercial DLS sizer, the lowest volume fraction in water yields a much better signal-to-noise ratio ($\approx 10^3$) than obtained with the highest volume fraction in oil ($\approx \num{90}$).
For the two lowest volume fractions in oil ($\phi=10^{-2}$ and $10^{-3}$), the SNR is too small with DLS ($\textrm{SNR}=10,1$ respectively), but CASS still provides the particle size with a low error (respectively 3 and 8 \% - Figure \ref{fig:S2}, Supplementary Materials).

\section{Discussion}

The key idea of the present work could be summarized as follows.
In our Lambertian cavities, light scattering probes the whole sample volume $V_{sa}$ multiple times, while classical DLS does it only once and only for the scatterers located inside the generally smaller volume illuminated by the laser light. 
The relevant size of the sample is no longer its length $l_{sa}$ along the light path, but the mean path length taken by light each time it travels across the sample, \textit{i.e.} the so-called mean chord length.
The eponym theorem states that the mean chord length distance for crossing the volume $V_{sa}$ enclosed by the surface area $\Sigma_{sa}$ reads $4V_{sa}/\Sigma_{sa}$ \cite{blanco2003,blanco2006short}, and the path length invariance \cite{savo2017observation} tells us that this mean distance is the same, regardless of the scattering process.
The mean sample path length therefore simply reads  $l_{sa}=4V_{sa}/\Sigma_{sa}$ (see Supplementary Materials), even in multiple scattering conditions.
As a consequence the total path length inside the sample once amplified is $g_{0} 4V_{sa}/\Sigma_{sa}$.
Because the cavity with volume $V_c$ and surface area $\Sigma_c$ is larger than the sample, photons traveling through the cavity have a probability $\Sigma_{sa}/\Sigma_c$ to probe the sample between two successive reflections on the walls.
However, the probability for one scatterer with scattering cross-section $\sigma_{sc}$ to scatter a random chord is simply $4\sigma_{sc}/\Sigma_{c}$.
The surface occupancy ratio $\sigma_{sc}/\Sigma_c$ indicates that the same scattering efficiency would be obtained if each scatterer would be projected on the cavity surface with an equivalent cross-section $2\sigma_{sc}$.
Incidentally, the same geometric argument applies to light absorption, and the probability for a single absorber with absorption cross-section $\sigma_a$ to intersect a single chord is proportional to the surface occupancy ratio $\sigma_{a}/\Sigma_c$. 
Using a calibrated solution of Crystal Violet dye, we found that absorbance could be measured down to $\SI{5e-5}{\per\centi\meter}$ in quartz cavities, and  $\SI{1e-3}{\per\centi\meter}$ with teflon cavities.

The concept of scattering amplification begs the question to know if the gain can be increased \textit{ad infinitum}. 
The very principle of DLS is based on fluctuating interferences, which require at least two scatterers to move with independent motions.
In the cavity however, unlike DLS, light can possibly be scattered multiple times by a single scatterer, leading to fluctuating interferences if that single scatterer alone is moving. 
This limit case is potentially interesting, but it is beyond the scope of our work and it points to a practical limit of CASS.
Namely, the phase variance is amplified proportionally to the enhancement factor given by equation \ref{CASS enhancement} only if individual scatterers contribute to at most one scattering event to the signal collected by the probe fiber, \textit{i.e.} if the probability of a scatterer to be sampled twice or more by the field entering the probe fiber remains negligible.
An excessive gain would destroy the linearity of the amplification of the phase variance. 
Even more interesting but worse, it potentially would possibly introduce closed loops in photon paths, that lead to large conductance fluctuations and additional interference effects \cite{Scheffold98_conductance}.
Entering this regime of "individual multiplicity" by single scatterers sets the limit to our scattering amplification strategy.

This linear amplification limit, or oversampling limit, obviously depends on how much light is collected by the probe fiber (\textit{i.e.} on the photon collection rate $f_{col}$) and how much time photons spend in the cavity on average (\textit{i.e.} the photon transit time $\tau_{tr}$), and these two parameters lead to the effective number of photons $\tilde{N}_{ph}=f_{col} \tau_{tr}$ that enter the average intensity.
Considering that the probability for a scatterer to contribute once to the signal is given by the surface occupancy ratio, amplified by $g_e$ and multiplied by $\tilde{N}_{ph}$, we can construct the sampling factor as 
$F=\tilde{N}_{ph}[g_{e}\sigma_{sc}/\Sigma_{c}]$, and the amplification limit simply reads $F^2\ll 1$.
In our experimental conditions, with a typical photon rate $f_{col}\approx \SI{5.1e5}{}$ photon.s$^{-1}$, with  transit time $\tau_{tr}\approx \SI{200}{\nano\second}$, and for $\SI{5}{\micro\meter}$ PMMA particles in water ($\sigma_{sc}\approx \SI{44}{\square\micro\meter}$), the sampling factor is such ($F\approx 6\times10^{-7}$) that the condition of linear amplification is safely met.
This means in principle that the SNR could be improved beyond our experimental conditions, by increasing the intensity collected the fiber by a maximal factor of $1/F\approx 1.7\times10^6$ before reaching oversampling.
This can be achieved by increasing the laser power injected into the cavity, but the detecting fiber must remain a single mode fiber to avoid field correlations that degrade intensity fluctuations. 
The aforementioned sampling factor $F$ also leads to a very simple way to re-scale the intensity decorrelation function $\mathcal{D(\tau)}$ shown on figure \ref{fig:3}. 
Out of $N_p$ particles in the sample, only $[N_pF]$ contribute to the decorrelation signal.
Using the definition of $\beta_{cass}$ in equation \ref{exponentiel beta CASS}, and for $\beta_{cass}\tau \ll 1$, we obtain the elementary decorrelation by the normalization $\mathcal{D}_{0}(\tau) = \mathcal{D}/N_p F$.
This simple derivation is the scheme we used to infer particle sizes from the intensity auto-correlation. 
In addition, knowing the decorrelation noise level  
$\sigma_{|g^{(1)}|^2}(\tau)$ of Fig.\ref{fig:4} (c), the condition $\textrm{SNR}\gg 1$ leads to a detection limit 
$\delta r_{min} (\tau)=\sqrt{N_p F  \sigma_{|g^{(1)}|^2} / [g_e p^{\star} k_{0}^{2}]}$.
Practically, in the conditions of figure \ref{fig:3}, 
\textit{i.e.} $N_p\approx 5.\num{e8}$, 
$l^{\star}\approx \SI{150}{\micro\meter}$, 
$F\approx \SI{9e-6}{}$, and 
$\lambda=2\pi/k_{0}=\SI{660}{\nano\meter}$, we find 
$\delta r_{min}  \approx \SI{200}{\pico\meter}$.

The CASS method presented here considerably extends the range of dynamic light scattering for samples that are too diluted, too small, or too transparent for classical DLS.
Major benefits can be envisioned for microfluidics, nephelometry for environmental control (water, air), or the characterization of various kinds of samples in the pharmaceutical industry. 
Beyond these applications based on detecting the sub-\r{a}ngstrom motions of scattering particles in gas or liquid phases, any bulk sample (including solid) that produces dynamic scattering due to density or dielectric fluctuations excited by thermal energy or external forcing could be studied.
The gain provided by CASS applies to these scattering fluctuations as much as they do for DLS, as shown by our recent marker-free observation of the internal dynamics of proteins \cite{graciani2021SPIE}, and new perspectives are opened for the development of more sensitive or miniature optical methods over a wide 
frequency range of 8 to 10 decades.

\section{Data availability}
The data that support the findings of this study are available from the corresponding author upon reasonable request.  

\begin{acknowledgement}

This work was supported by the taxpayers of South Korea through the Institute for Basic Science, Project Code IBS-R020-D1.

\end{acknowledgement}

\section{Author contributions}
F.A directed the project, G.G performed the experiments and analyzed the data, G.G, F.A. and J.T.K. wrote the manuscript.

\section{Competing interests}
The Authors declare no competing interests

\bibliography{main.bib}

\providecommand{\latin}[1]{#1}
\makeatletter
\providecommand{\doi}
  {\begingroup\let\do\@makeother\dospecials
  \catcode`\{=1 \catcode`\}=2 \doi@aux}
\providecommand{\doi@aux}[1]{\endgroup\texttt{#1}}
\makeatother
\providecommand*\mcitethebibliography{\thebibliography}
\csname @ifundefined\endcsname{endmcitethebibliography}
  {\let\endmcitethebibliography\endthebibliography}{}
\begin{mcitethebibliography}{28}
\providecommand*\natexlab[1]{#1}
\providecommand*\mciteSetBstSublistMode[1]{}
\providecommand*\mciteSetBstMaxWidthForm[2]{}
\providecommand*\mciteBstWouldAddEndPuncttrue
  {\def\EndOfBibitem{\unskip.}}
\providecommand*\mciteBstWouldAddEndPunctfalse
  {\let\EndOfBibitem\relax}
\providecommand*\mciteSetBstMidEndSepPunct[3]{}
\providecommand*\mciteSetBstSublistLabelBeginEnd[3]{}
\providecommand*\EndOfBibitem{}
\mciteSetBstSublistMode{f}
\mciteSetBstMaxWidthForm{subitem}{(\alph{mcitesubitemcount})}
\mciteSetBstSublistLabelBeginEnd
  {\mcitemaxwidthsubitemform\space}
  {\relax}
  {\relax}

\bibitem[Berne and Pecora(2000)Berne, and Pecora]{berne2000dynamic}
Berne,~B.~J.; Pecora,~R. \emph{Dynamic light scattering: with applications to
  chemistry, biology, and physics}; Courier Corporation, 2000\relax
\mciteBstWouldAddEndPuncttrue
\mciteSetBstMidEndSepPunct{\mcitedefaultmidpunct}
{\mcitedefaultendpunct}{\mcitedefaultseppunct}\relax
\EndOfBibitem
\bibitem[Stokes()]{stokes8g}
Stokes,~C. G. 1845 On the theories of the internal friction of fluids in
  motion, and of the equilibrium and motion of elastic solids. \emph{Trans.
  Camb. phil. Soc. 8, 287--341 (reprinted in Mathematical and physical papers}
  \emph{1}, 75--129\relax
\mciteBstWouldAddEndPuncttrue
\mciteSetBstMidEndSepPunct{\mcitedefaultmidpunct}
{\mcitedefaultendpunct}{\mcitedefaultseppunct}\relax
\EndOfBibitem
\bibitem[Einstein(1905)]{einstein1905erzeugung}
Einstein,~A. {\"U}ber einen die Erzeugung und Verwandlung des Lichtes
  betreffenden heuristischen Gesichtspunkt. \emph{Annalen der physik}
  \textbf{1905}, \emph{322}, 132--148\relax
\mciteBstWouldAddEndPuncttrue
\mciteSetBstMidEndSepPunct{\mcitedefaultmidpunct}
{\mcitedefaultendpunct}{\mcitedefaultseppunct}\relax
\EndOfBibitem
\bibitem[Einstein(1906)]{einstein1906theorie}
Einstein,~A. Zur theorie der brownschen bewegung. \emph{Annalen der physik}
  \textbf{1906}, \emph{324}, 371--381\relax
\mciteBstWouldAddEndPuncttrue
\mciteSetBstMidEndSepPunct{\mcitedefaultmidpunct}
{\mcitedefaultendpunct}{\mcitedefaultseppunct}\relax
\EndOfBibitem
\bibitem[Einstein(1910)]{einstein1910theory}
Einstein,~A. Theory of opalescence of homogenous liquids and liquid mixtures
  near critical conditions. \emph{Annalen Der Physik} \textbf{1910}, \emph{33},
  1275--1298\relax
\mciteBstWouldAddEndPuncttrue
\mciteSetBstMidEndSepPunct{\mcitedefaultmidpunct}
{\mcitedefaultendpunct}{\mcitedefaultseppunct}\relax
\EndOfBibitem
\bibitem[Maret and Wolf(1987)Maret, and Wolf]{maret1987}
Maret,~G.; Wolf,~P. Multiple light scattering from disordered media: the effect
  of Brownian motion of scatterers. \emph{Z Phys} \textbf{1987}, \emph{65},
  409--413\relax
\mciteBstWouldAddEndPuncttrue
\mciteSetBstMidEndSepPunct{\mcitedefaultmidpunct}
{\mcitedefaultendpunct}{\mcitedefaultseppunct}\relax
\EndOfBibitem
\bibitem[Pine \latin{et~al.}(1988)Pine, Weitz, Chaikin, and
  Herbolzheimer]{pine1988diffusing}
Pine,~D.; Weitz,~D.; Chaikin,~P.; Herbolzheimer,~E. Diffusing wave
  spectroscopy. \emph{Physical review letters} \textbf{1988}, \emph{60},
  1134\relax
\mciteBstWouldAddEndPuncttrue
\mciteSetBstMidEndSepPunct{\mcitedefaultmidpunct}
{\mcitedefaultendpunct}{\mcitedefaultseppunct}\relax
\EndOfBibitem
\bibitem[MacKintosh and John(1989)MacKintosh, and
  John]{mackintosh1989diffusing}
MacKintosh,~F.; John,~S. Diffusing-wave spectroscopy and multiple scattering of
  light in correlated random media. \emph{Physical Review B} \textbf{1989},
  \emph{40}, 2383\relax
\mciteBstWouldAddEndPuncttrue
\mciteSetBstMidEndSepPunct{\mcitedefaultmidpunct}
{\mcitedefaultendpunct}{\mcitedefaultseppunct}\relax
\EndOfBibitem
\bibitem[Scheffold \latin{et~al.}(2001)Scheffold, Skipetrov, Romer, and
  Schurtenberger]{scheffold2001diffusing}
Scheffold,~F.; Skipetrov,~S.; Romer,~S.; Schurtenberger,~P. Diffusing-wave
  spectroscopy of nonergodic media. \emph{Physical Review E} \textbf{2001},
  \emph{63}, 061404\relax
\mciteBstWouldAddEndPuncttrue
\mciteSetBstMidEndSepPunct{\mcitedefaultmidpunct}
{\mcitedefaultendpunct}{\mcitedefaultseppunct}\relax
\EndOfBibitem
\bibitem[Zakharov and Scheffold(2009)Zakharov, and
  Scheffold]{zakharov2009advances}
Zakharov,~P.; Scheffold,~F. \emph{Light Scattering Reviews 4}; Springer, 2009;
  pp 433--467\relax
\mciteBstWouldAddEndPuncttrue
\mciteSetBstMidEndSepPunct{\mcitedefaultmidpunct}
{\mcitedefaultendpunct}{\mcitedefaultseppunct}\relax
\EndOfBibitem
\bibitem[Kim \latin{et~al.}(2019)Kim, {\c{S}}enbil, Zhang, Scheffold, and
  Mason]{kim2019diffusing}
Kim,~H.~S.; {\c{S}}enbil,~N.; Zhang,~C.; Scheffold,~F.; Mason,~T.~G. Diffusing
  wave microrheology of highly scattering concentrated monodisperse emulsions.
  \emph{Proceedings of the National Academy of Sciences} \textbf{2019},
  201817029\relax
\mciteBstWouldAddEndPuncttrue
\mciteSetBstMidEndSepPunct{\mcitedefaultmidpunct}
{\mcitedefaultendpunct}{\mcitedefaultseppunct}\relax
\EndOfBibitem
\bibitem[Le~Goff \latin{et~al.}(2001)Le~Goff, Amblard, and Furst]{le2001motor}
Le~Goff,~L.; Amblard,~F.; Furst,~E.~M. Motor-driven dynamics in actin-myosin
  networks. \emph{Physical review letters} \textbf{2001}, \emph{88},
  018101\relax
\mciteBstWouldAddEndPuncttrue
\mciteSetBstMidEndSepPunct{\mcitedefaultmidpunct}
{\mcitedefaultendpunct}{\mcitedefaultseppunct}\relax
\EndOfBibitem
\bibitem[Graciani \latin{et~al.}(2020)Graciani, Le~Goff, and
  Amblard]{graciani2020protein}
Graciani,~G.; Le~Goff,~L.; Amblard,~F. Protein Conformational Dynamics Probed
  Correlation Spectroscopy of Multiply Scattered Light. \emph{Biophysical
  Journal} \textbf{2020}, \emph{118}, 136a\relax
\mciteBstWouldAddEndPuncttrue
\mciteSetBstMidEndSepPunct{\mcitedefaultmidpunct}
{\mcitedefaultendpunct}{\mcitedefaultseppunct}\relax
\EndOfBibitem
\bibitem[Ulbricht(1905)]{ulbricht1905}
Ulbricht,~R. \emph{Die Vorg{\"a}nge im Kugelphotometer}; 1905\relax
\mciteBstWouldAddEndPuncttrue
\mciteSetBstMidEndSepPunct{\mcitedefaultmidpunct}
{\mcitedefaultendpunct}{\mcitedefaultseppunct}\relax
\EndOfBibitem
\bibitem[Graciani \latin{et~al.}(2021)Graciani, Filoche, and
  Amblard]{graciani20213d}
Graciani,~G.; Filoche,~M.; Amblard,~F. 3D stochastic interferometer detects
  picometer deformations and minute dielectric fluctuations of its optical
  volume. \emph{arXiV:2110.07390} \textbf{2021}, \relax
\mciteBstWouldAddEndPunctfalse
\mciteSetBstMidEndSepPunct{\mcitedefaultmidpunct}
{}{\mcitedefaultseppunct}\relax
\EndOfBibitem
\bibitem[Graciani and Amblard(2019)Graciani, and Amblard]{graciani2019random}
Graciani,~G.; Amblard,~F. Random dynamic interferometer: cavity amplified
  speckle spectroscopy using a highly symmetric coherent field created inside a
  closed Lambertian optical cavity. \emph{Applied Optical Metrology III}
  \textbf{2019}, \emph{11102}, 111020N\relax
\mciteBstWouldAddEndPuncttrue
\mciteSetBstMidEndSepPunct{\mcitedefaultmidpunct}
{\mcitedefaultendpunct}{\mcitedefaultseppunct}\relax
\EndOfBibitem
\bibitem[Cone \latin{et~al.}(2015)Cone, Musser, Figueroa, Mason, and
  Fry]{cone2015diffuse}
Cone,~M.~T.; Musser,~J.~A.; Figueroa,~E.; Mason,~J.~D.; Fry,~E.~S. Diffuse
  reflecting material for integrating cavity spectroscopy, including ring-down
  spectroscopy. \emph{Applied optics} \textbf{2015}, \emph{54}, 334--346\relax
\mciteBstWouldAddEndPuncttrue
\mciteSetBstMidEndSepPunct{\mcitedefaultmidpunct}
{\mcitedefaultendpunct}{\mcitedefaultseppunct}\relax
\EndOfBibitem
\bibitem[Fry \latin{et~al.}(2006)Fry, Musser, Kattawar, and
  Zhai]{fry2006integrating}
Fry,~E.~S.; Musser,~J.; Kattawar,~G.~W.; Zhai,~P.-W. Integrating cavities:
  temporal response. \emph{Applied optics} \textbf{2006}, \emph{45},
  9053--9065\relax
\mciteBstWouldAddEndPuncttrue
\mciteSetBstMidEndSepPunct{\mcitedefaultmidpunct}
{\mcitedefaultendpunct}{\mcitedefaultseppunct}\relax
\EndOfBibitem
\bibitem[Fry and Mason(2016)Fry, and Mason]{fry2016integratingicrds}
Fry,~E.~S.; Mason,~J.~D. Integrating cavity ring-down spectroscopy (ICRDS) and
  the direct measurement of absorption coefficients. \emph{Physica Scripta}
  \textbf{2016}, \emph{91}, 043004\relax
\mciteBstWouldAddEndPuncttrue
\mciteSetBstMidEndSepPunct{\mcitedefaultmidpunct}
{\mcitedefaultendpunct}{\mcitedefaultseppunct}\relax
\EndOfBibitem
\bibitem[Khaoua \latin{et~al.}(2021)Khaoua, Graciani, Kim, and
  Amblard]{khaoua2021stochastic}
Khaoua,~I.; Graciani,~G.; Kim,~A.; Amblard,~F. Stochastic light concentration
  from 3D to 2D reveals ultraweak chemi-and bioluminescence. \emph{Scientific
  reports} \textbf{2021}, \emph{11}, 1--13\relax
\mciteBstWouldAddEndPuncttrue
\mciteSetBstMidEndSepPunct{\mcitedefaultmidpunct}
{\mcitedefaultendpunct}{\mcitedefaultseppunct}\relax
\EndOfBibitem
\bibitem[Fry \latin{et~al.}(1992)Fry, Kattawar, and Pope]{fry1992}
Fry,~E.~S.; Kattawar,~G.~W.; Pope,~R.~M. Integrating cavity absorption meter.
  \emph{Applied Optics} \textbf{1992}, \emph{31}, 2055--2065\relax
\mciteBstWouldAddEndPuncttrue
\mciteSetBstMidEndSepPunct{\mcitedefaultmidpunct}
{\mcitedefaultendpunct}{\mcitedefaultseppunct}\relax
\EndOfBibitem
\bibitem[Ju \latin{et~al.}(1992)Ju, Frank, and Gast]{ju1992contin}
Ju,~R.~T.; Frank,~C.~W.; Gast,~A.~P. CONTIN analysis of colloidal aggregates.
  \emph{Langmuir} \textbf{1992}, \emph{8}, 2165--2171\relax
\mciteBstWouldAddEndPuncttrue
\mciteSetBstMidEndSepPunct{\mcitedefaultmidpunct}
{\mcitedefaultendpunct}{\mcitedefaultseppunct}\relax
\EndOfBibitem
\bibitem[S.~Blanco(2003)]{blanco2003}
S.~Blanco,~R.~F. An invariance property of diffusive random walks.
  \emph{Europhys. Lett.} \textbf{2003}, \emph{61}, 168--173\relax
\mciteBstWouldAddEndPuncttrue
\mciteSetBstMidEndSepPunct{\mcitedefaultmidpunct}
{\mcitedefaultendpunct}{\mcitedefaultseppunct}\relax
\EndOfBibitem
\bibitem[Blanco and Fournier(2006)Blanco, and Fournier]{blanco2006short}
Blanco,~S.; Fournier,~R. Short-path statistics and the diffusion approximation.
  \emph{Physical review letters} \textbf{2006}, \emph{97}, 230604\relax
\mciteBstWouldAddEndPuncttrue
\mciteSetBstMidEndSepPunct{\mcitedefaultmidpunct}
{\mcitedefaultendpunct}{\mcitedefaultseppunct}\relax
\EndOfBibitem
\bibitem[Savo \latin{et~al.}(2017)Savo, Pierrat, Najar, Carminati, Rotter, and
  Gigan]{savo2017observation}
Savo,~R.; Pierrat,~R.; Najar,~U.; Carminati,~R.; Rotter,~S.; Gigan,~S.
  Observation of mean path length invariance in light-scattering media.
  \emph{Science} \textbf{2017}, \emph{358}, 765--768\relax
\mciteBstWouldAddEndPuncttrue
\mciteSetBstMidEndSepPunct{\mcitedefaultmidpunct}
{\mcitedefaultendpunct}{\mcitedefaultseppunct}\relax
\EndOfBibitem
\bibitem[Scheffold and Maret(1998)Scheffold, and
  Maret]{Scheffold98_conductance}
Scheffold,~F.; Maret,~G. Universal Conductance Fluctuations of Light.
  \emph{Phys. Rev. Lett.} \textbf{1998}, \emph{81}, 5800--5803\relax
\mciteBstWouldAddEndPuncttrue
\mciteSetBstMidEndSepPunct{\mcitedefaultmidpunct}
{\mcitedefaultendpunct}{\mcitedefaultseppunct}\relax
\EndOfBibitem
\bibitem[Graciani \latin{et~al.}()Graciani, King, and
  Amblard]{graciani2021SPIE}
Graciani,~G.; King,~J.~T.; Amblard,~F. Marker-free protein study by amplified
  light scattering. \emph{SPIE Proceedings, under revision} \relax
\mciteBstWouldAddEndPunctfalse
\mciteSetBstMidEndSepPunct{\mcitedefaultmidpunct}
{}{\mcitedefaultseppunct}\relax
\EndOfBibitem
\end{mcitethebibliography}

\clearpage
\addtocounter{figure}{-1}
\begin{figure} [t!]
[Figure 1 Caption] 
Principle of cavity-amplified scattering spectroscopy (CASS). 
(a) Experimental setup of CASS. (b-c) Ray-tracing comparison of light paths for Dynamic Light Scattering (DLS) and Diffusing Wave Spectroscopy (DWS) (b) and for Cavity Amplified Scattering Spectroscopy (c), when a plane wave of wavevector $\textbf{k}_{0}$ enters a sample.
In DLS, the intensity resulting from the superposition of multiple paths (here n=2) is collected at times t and $t+\tau$ in the direction $\textbf{u}=\textbf{k}_{s}/k_0$, with scattering vector $\textbf{q}=\textbf{k}_{s}-\textbf{k}_{0}$. 
In CASS, scattered light is measured with a single mode fiber that collects light over its whole angular aperture.
Each scatterer moves by an amplitude $\delta \textbf{r}$ during $\tau$. 
\end{figure}
\thispagestyle{empty}
\clearpage
\addtocounter{figure}{+1}
\begin{figure}
\includegraphics[width=\linewidth]{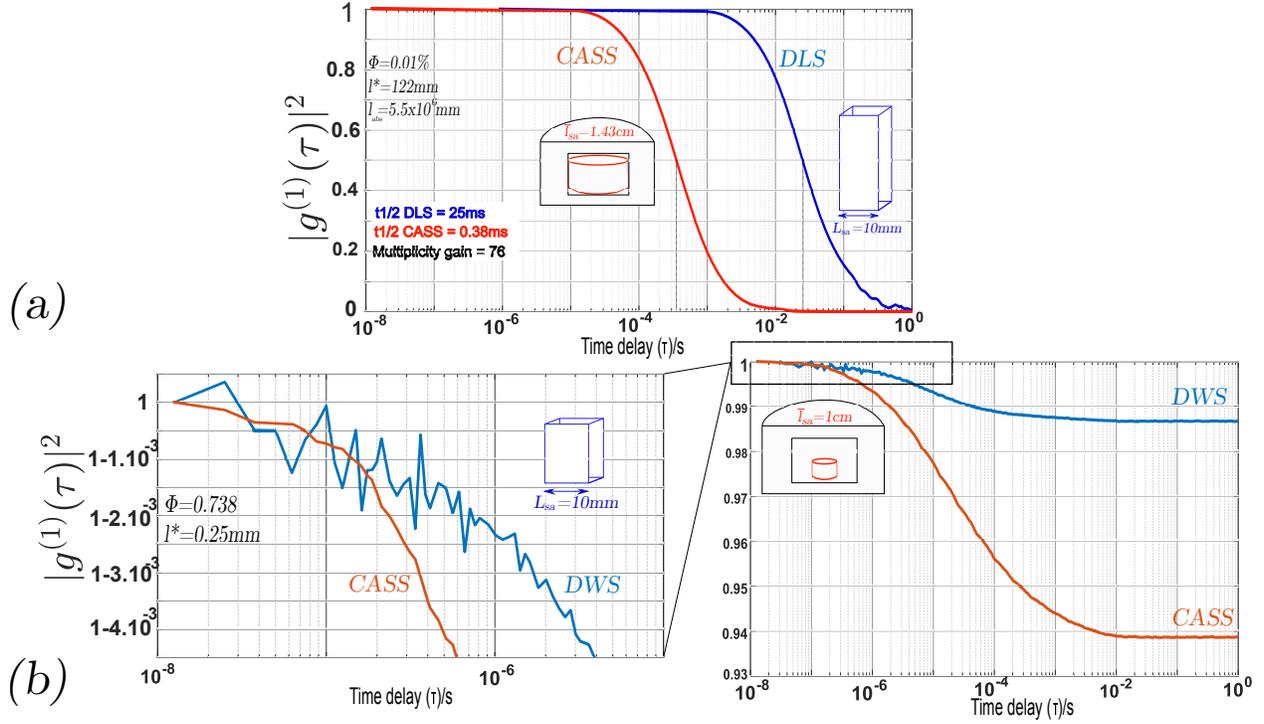}
\caption{\label{fig:2} Effect of cavity amplification on DLS and DWS. (a) Normalized light intensity auto-correlation $|g^{(1)}(\tau)|^{2}$ of a suspension of $\SI{5.5}{\micro\meter}$ diameter PMMA particles in water obtained by DLS (blue line) and cavity-amplified DLS (CASS-DLS, red line). 
$t_{1/2}$ is defined as the half-decorrelation time. 
(b, right) Normalized light intensity auto-correlation $|g^{(1)}(\tau)|^{2}$ of a SDS-stabilized mono disperse oil/water emulsion \cite{kim2019diffusing} using Conventional DWS (blue) and cavity-amplified DWS (CASS-DWS, red). 
(a, left) Zoom on short time delays from $\SI{12.5}{\nano\second}$ to $\SI{10}{\micro\second}$.}
\end{figure}
\clearpage
\begin{figure}
\includegraphics[width=\linewidth]{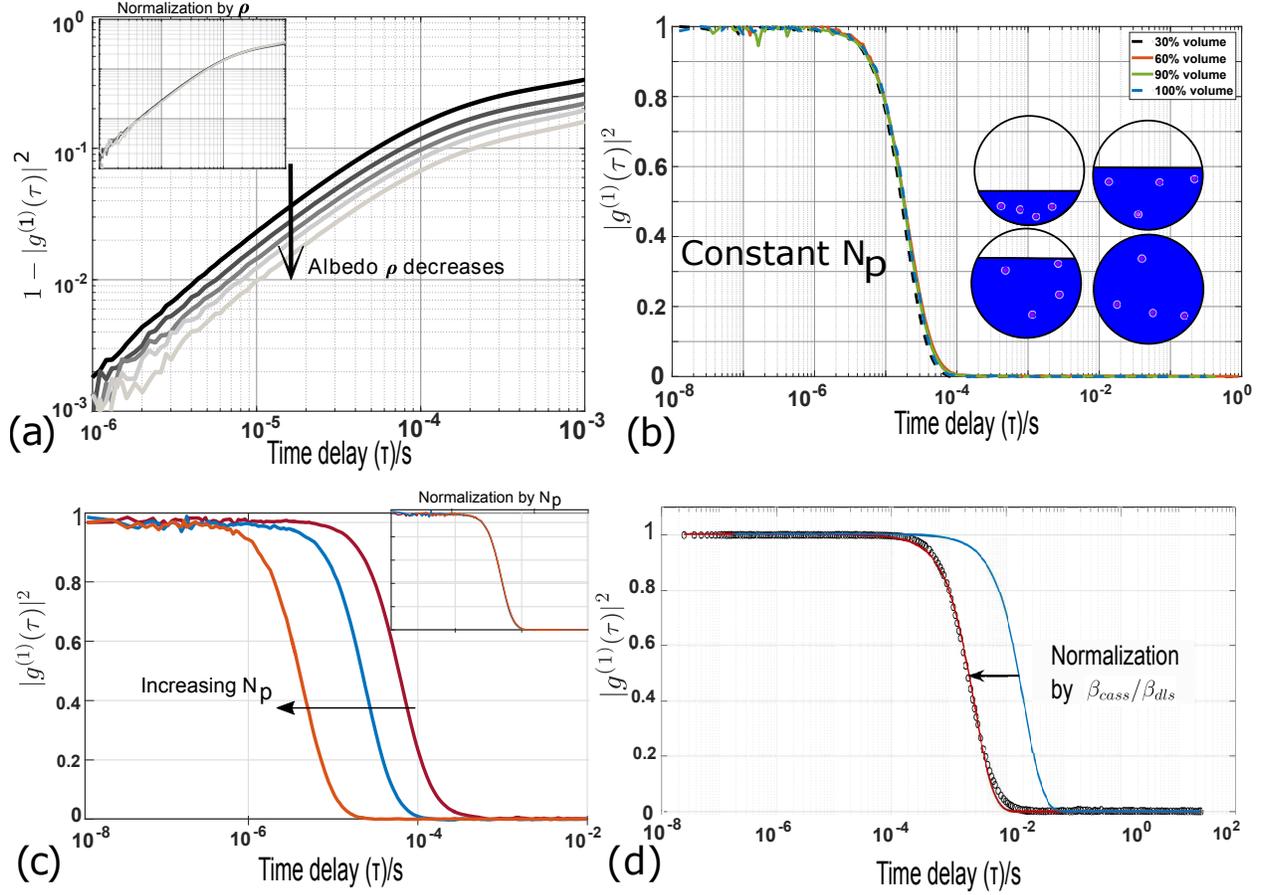}
\caption{\label{fig:3} Properties of the cavity amplification. 
(a) Normalized light intensity decorrelation ($\mathcal{D}(\tau)=1-|g^{(1)}(\tau)|^{2}$) induced by the thermal fluctuations of a suspension of $\SI{27}{\nano\meter}$ diameter polystyrene particles in a Cavity Amplified Scattering Spectroscopy (CASS) setup. 
The albedo is incrementally decreased from top to bottom (\SIlist{.991;.988; .985;.983;.978}{}). 
The insert on top shows the same graphs with normalization by the cavity albedo $\rho$.
(b) Normalized light intensity auto-correlation ($|g^{(1)}(\tau)|^{2}$) from a suspension of $\SI{5.5}{\micro\meter}$ PMMA in different volume of water in a commercial Spectralon cavity. 
The same number of particles was used in an increasing volume of water, filling $30\%$, $60\%$, $90\%$ and $100\%$ of the total cavity volume.
(c) Normalized light intensity auto-correlation 
($|g^{(1)}(\tau)|^{2}$) from a suspension of $\SI{5.5}{\micro\meter}$ PMMA in water in a teflon cavity. 
From right to left, the number of particles is successively increased 10 and 100 folds. 
The insert on the right shows the same graphs after normalization by the number of particles.
(d) Normalized light intensity auto-correlation from a suspension of $\SI{20}{\micro\meter}$ diameter PMMA particles in water: 
experimental raw data for a CASS measurement in a teflon cavity 
(black circles), 
numerical simulation for a non-amplified DLS measurement (blue line), 
and simulation corrected for by the enhancement ratio  $\beta_{cass}/\beta_{dls}^{(\theta=\pi/2)}$ (red line).}
\end{figure}

\clearpage
\begin{figure*}
\includegraphics[width=\linewidth]{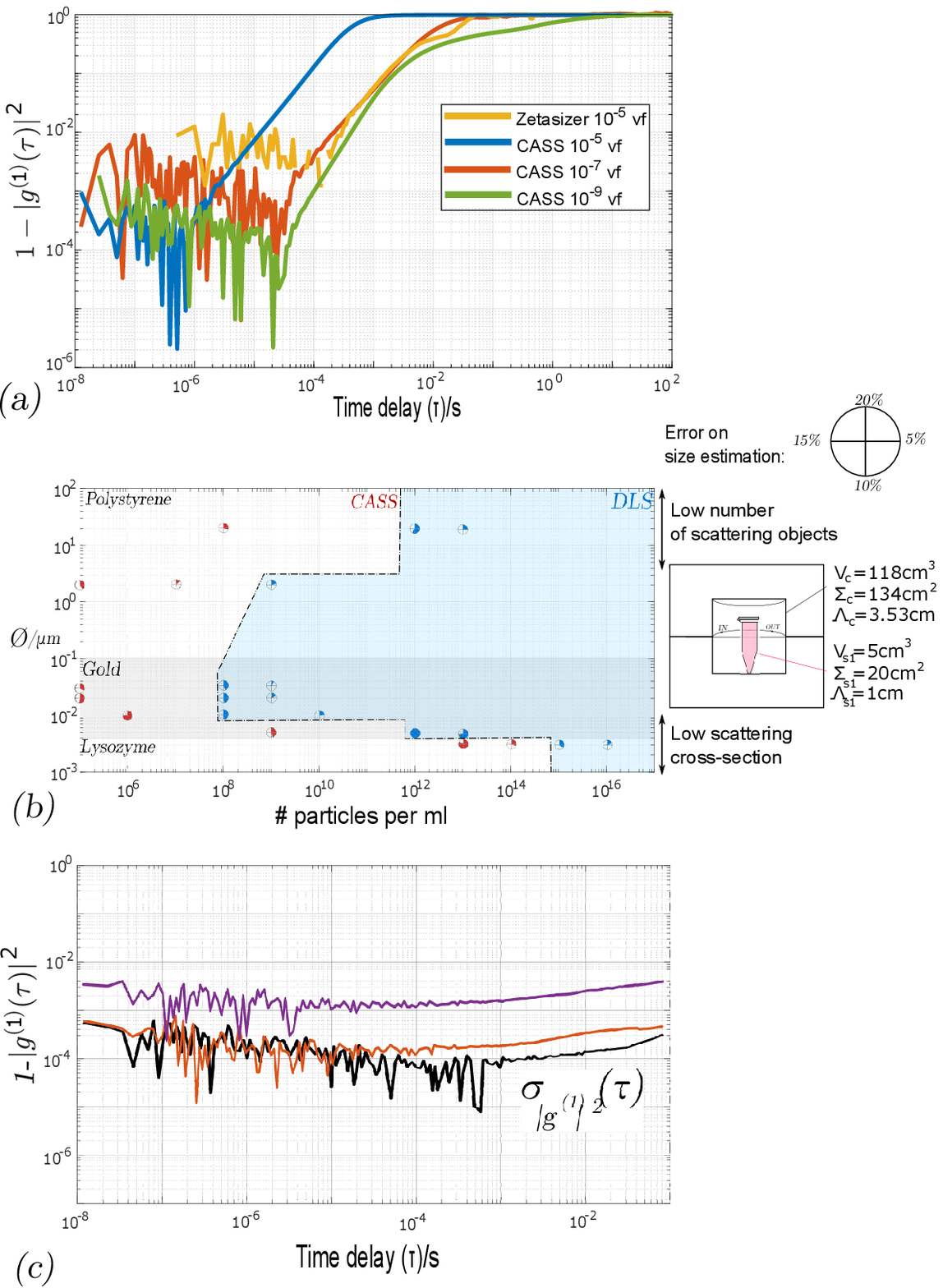}
\caption{\label{fig:4}}
\end{figure*}
\addtocounter{figure}{-1}
\begin{figure} [t!]
[Figure 4 Caption] 
 Performance of cavity amplification.
(a) Comparison of decorrelation functions $1-|g^{(1)}(\tau)|^{2}$ between the commercial DLS instrument Zetasizer and our quartz CASS setup for $\SI{20}{\micro\meter}$ polystyrene particles, and various volume fractions: $10^{-5}$ for DLS (yellow line) and $[10^{-5},10^{-7},10^{-9}]$ for CASS (resp. blue, red and green).
(b, left) Comparison of the sensitivity of DLS and CASS for different concentrations of different scattering objects: 
Polystyrene particles ($\SI{20}{\micro\meter}$ and $\SI{2}{\micro\meter}$ diameter), 
Gold particles ($\SI{30}{}$, $\SI{20}{}$, $\SI{10}{}$ and $\SI{5}{\nano\meter}$ diameter), 
and a protein (lysozyme, 15kDa molecular mass)).
Particle sizes were measured by DLS (blue disks) and CASS (red disks) with an error represented by the colored fraction of each disk. 
For each sample type, the Zetasizer reaches a sensitivity limit indicated by a minimal concentration (black dotted line).
(b, right) schematic setup for CASS measurements. 
(c) CASS response for $\SI{30}{\nano\meter}$ Gold nanoparticles at the detection limit ($10^5$ particles per ml, red) and for 10-folds higher concentration (purple). 
The black line represents the relevant noise in these conditions, \textit{i.e.}
the standard deviation of  $|g^{(1)}(\tau)|^{2}$.
\end{figure}\thispagestyle{empty}

\clearpage
\addtocounter{figure}{+1}
\begin{figure*}
\includegraphics[width=\linewidth]{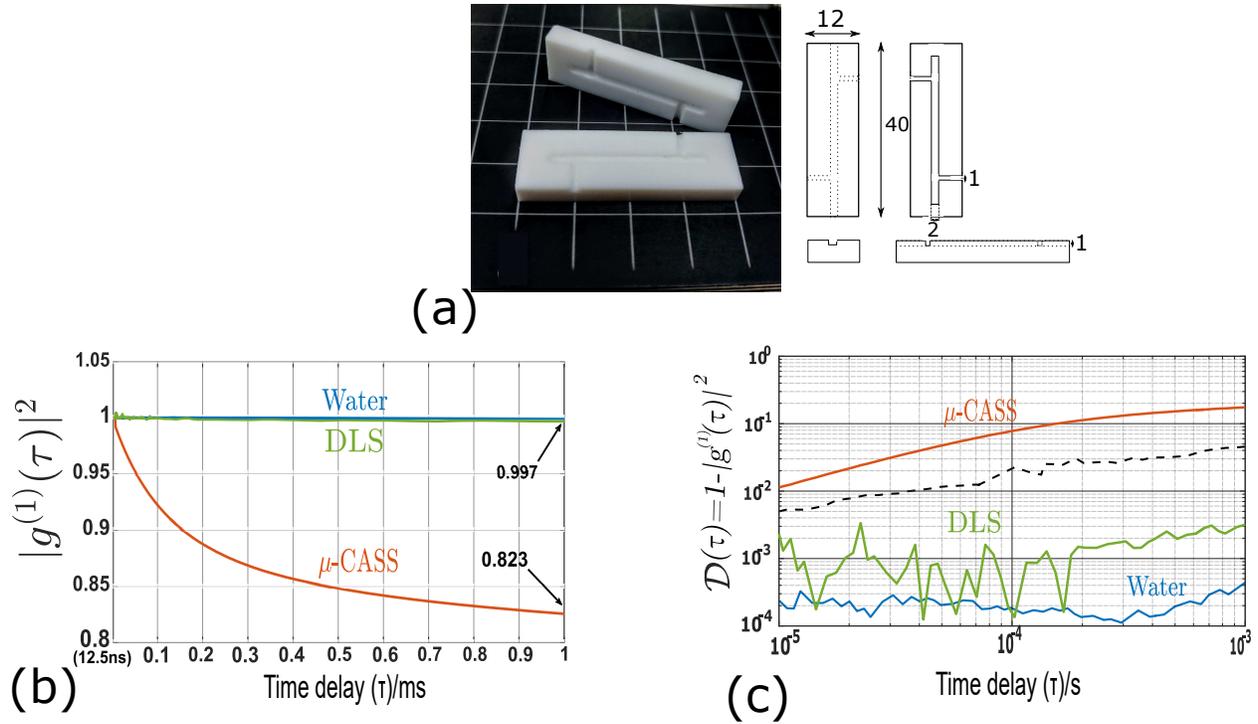}
\caption{\label{fig:5}
Response of miniature CASS.
(a) Picture and drawing of a \SI{100}{\micro\liter} capacity teflon cavity with its input/output fiber ports and sample chamber (white lines show the scale, with \SI{10}{\milli\meter} between them).\\
(b-c) Intensity auto-correlation $|g^{(1)}(\tau)^{2}|$ of a \SI{100}{\\micro\liter} suspension of \SI{27}{\nano\meter} diameter polystyrene particles in water in the miniaturized cavity (red), pure water (blue) and conventional DLS (green), with \SI{5}{\min} acquisition time.
Linear auto-correlation curves (b) and log-log decorrelation curves (c) are shown together with the the noise level of CASS (c,black dotted line).}
\end{figure*}
\clearpage
\newpage
\section{Methods}

\subsubsection{Light scattering experiments} 

A \SI{660}{\nano\meter} single frequency (linewidth $<\SI{1}{\mega\hertz}$) \SI{300}{\milli\watt} continuous wave laser (Cobolt Flamenco 04-01) was used for all light scattering experiments. 
Light was collected with a split monomode optical fiber ($\SI{670}{}\pm\SI{50}{\nano\meter}$, $<\SI{3.9}{\deci\bel}$ insertion loss), and sent to two avalanche single photon counting modules (SPCM-AQRH, Excelitas) with: \SI{180}{\micro\meter} diameter, $65\%$ efficiency @ \SI{650}{\nano\meter}, \SI{10}{\nano\second} output pulse width, \SI{22}{\nano\second} dead time, $1.4\%$ afterpulsing probability.
Intensity auto-correlation was measured from the cross-correlation of the two photodiodes to eliminate afterpulsing, using a digital correlator from LSI Instruments, with: 
16/8 multi-tau correlation scheme, lag times spanning form \SI{12.5}{\nano\second} to \SI{3436}{\second}, 322 channels, lag time range \SI{54976}{\second}, maximum count rate \SI{20}{\mega\hertz} during \SI{52}{\milli\second} and over 2 channels. 
A 5.3 inches commercial integrating sphere (Newport, 819C-IS-5.3) was used as the amplifying cavity to obtain the results shown in Fig.\ref{fig:4} (a).
Home-made cavities machined in teflon and compressed fumed quartz powder were used otherwise. 
Their albedo were measured by assessing the response of the cavities to a \SI{670}{\nano\meter} laser (PicoQuant, PC-670M and PDL-800B driver) delivering $\tau_{\textrm{FWHM}}< \SI{120}{\pico\second}$ and $\SI{2}{\nano\joule}$ pulses, using a Time correlated single photon counting system (PicoQuant, PicoHarp 300 + PicoQuant SIA 400 attenuator/inverter module) with: 
a \SI{4}{\pico\second} time bin width, a dead time $<\SI{95}{\nano\second}$, and 2 channels with 16 bits. 
A Zetasizer Nano ZSP (Malvern) was used as the reference commercial DLS particle sizer instrument.

\subsubsection{Microparticles} 
Water and oil suspensions were prepared DLS and CASS experiments with $\SI{5.5}{\micro\meter}$ diameter Poly(methyl methacrylate) (PMMA) particles (Bangs Laboratories, catalog number BB01N, Inv$\#$ L090702B, Lot$\#$ 9285) and with $\SI{27}{\nano\meter}$ diameter  polystyrene  particles (Bangs Laboratories, catalog number PS02001, PS02N, Inv$\#$ L141204B, Lot$\#$ 11848). 
The oil-in-water emulsions used in Fig.\ref{fig:2} are mono-disperse polydimethyl siloxane (PDMS) droplets with average radius of \SI{459}{\nano\meter} kindly provided by Prof. Thomas G. Mason (mason@physics.ucla.edu, University of California, Los Angeles) who characterized their rms motion spectra \cite{kim2019diffusing}. 
High dilution experiments were made with Gold nanoparticles with diameters 5, 10, 20 and \SI{30}{\nano\meter} (Sigma Aldrich, respective catalog numbers 752568, 752584, 753610 and 753629).
Protein solutions were prepared for DLS and CASS experiments with lysozyme (Thermo Fisher Scientific, catalog 89833, lot 280426) in a Tris-EDTA buffer (pH 7.4).

\newpage
\section{Supplementary materials}



\subsubsection{\label{sec:level3} Path length statistics in an empty Lambertian cavity}

In the empty Lambertian cavity, a photon path $ s_{i}$ can be represented as a succession of $N_{ch}^{(i)}$ chords 
$Ch^{(i)}_{n=1,\ldots,N_{ch}^{(i)}}$ crossing the cavity alternating with diffuse reflection paths inside the cavity walls.
We can safely assume that the light paths through the walls are static and make no contribution to the light decorrelation.
As a consequence, and for the sake of simplicity, we essentially ignore them, under the provision that their contribution to the total path length remains negligible compared to chords, and/or that the total path length remains smaller than the coherence length.
Each path can then be written as 
$ s_{i} = \bigcup_{n=1}^{N_{ch}^{(i)}}Ch_{n}^{(i)}$, 
where the number of chords is a random variable with the probability distribution 
$p(N_{ch}) = \rho^{N_{ch}-1} (1-\rho)$.
The mean number of chords $\overline{N}_{ch}=-1/\ln{(\rho)}$ is identified as the reflection gain $g_{0}\approx 1/\epsilon$, determined by the wall albedo $ \rho $ or the reflection loss coefficient $ \epsilon = 1-\rho $.
If $l_{Ch}$ is the random variable that describes the length of individual chords and $\overline{l}_{Ch}$ its mean, the total length of a path $s_i$ is described by the random sum of identically and independently repeated chord lengths as 
$l_{c} = \sum _{ n=1 }^{ N_{c} }{ l_{Ch_{n}} }$, and the average length of all Lambertian paths simply reads 
$L_{c}=\overline{l}_{c} = g_{0} \overline{l}_{Ch}$. 
From the mean chord length theorem \cite{blanco2003}\cite{blanco2006short}, we know that the average chord length is given by the ratio $\overline{l}_{Ch} = 4 \frac{V_{c}}{\Sigma_{c}}$ of the cavity volume $V_{c}$ to the surface area $\Sigma_{c}$ of its inside walls, and the average length of Lambertian paths in the empty cavity therefore reads:
\begin{equation}
\overline{l}_{c} = \frac{4 g_{0} V_{c}}{\Sigma_{c}}
\end{equation}

\subsubsection{\label{sec:level2} Path length invariance, absorption losses, and mean total path through a sample}

Let's now consider that the cavity is totally filled with a scattering sample characterized by a mean transport length $ l^{\star}$ and a photon absorption length $l_{abs}$. 
While the albedo $\rho=1-\epsilon$ accounts for the probability to be reflected by the cavity walls, the probability of a photon to be absorbed by the sample between two successive reflections can be taken into account by considering an average absorption loss $\overline{\epsilon_{a}}$ through the sample, that leads to an effective albedo $\rho_{e}=\rho (1-\overline{\epsilon_{a}})$.
While the absorption loss $\epsilon_{ch}$ along a straight chord of length $l_{ch}$ through the sample simply reads 
$1-e^{\frac{-l_{ch}}{l_{abs}}}$, the absorption loss along a scattered or Brownian path necessarily depends on its particular length, which is random. 
However, because the mean chord length theorem leads to the mean path length invariance, the average length of scattering excursions through a 3D volume enclosed by a closed surface  does not depend of the scattering length \cite{savo2017observation}, and it is simply equal to the mean chord length $\overline{l}_{ch}$.
In other words, within the linear approximation
$\left<e^{-x}\right>=e^{-\left<x\right>}$ which holds for $x\ll1$ and is obviously needed for the amplification process, the average absorption loss over all possible scattered paths is merely derived from the average chord length, and the effective albedo reads 
$\rho_e= \rho \, e^{\frac{-\overline{l}_{ch}}{l_{abs}}}  \approx (1-\epsilon) (1-\frac{\overline{l}_{ch}}{l_{abs}})$.
We obtain an effective loss coefficient 
$ \epsilon_{e} = 1- \rho_{e} \approx \epsilon + \frac{\overline{l}_{ch}}{l_{abs}}$, 
that exceeds the reflection loss coefficient $\epsilon$ by the factor 
$R_a = 1 + \frac{\overline{l}_{ch}}{\epsilon l_{abs}}$. 
Obviously, the gain $g_0=1/\epsilon$ of the empty cavity must also be replaced by the effective gain $g_e = g_0/R_a$. 
The rather simple interpretation is that sample absorption can be neglected if 
$l_{abs} \gg g_0 \overline{l}_{ch}$. 
If the cavity is only partially filled, with a sample volume $ V_{sa}\leq V_c$ and surface $\Sigma_{sa}\leq \Sigma_c$, the average loss $\overline{\epsilon_{a}}$ experienced by Lambertian chords crossing the sample is simply derived from the average sample chord 
$\overline{l}_{sa} = \frac{4 V_{sa}}{\Sigma_{sa}}$, 
and the correction factor leading to the effective loss coefficient $R_a \epsilon$ and effective gain $g_0 / R_a$ now reads $R_a=1 + \frac{4V_{sa}}{\epsilon l_{abs} \Sigma_{sa}}$.

\subsubsection{Additional measurements for particle sizing}

Additional measurements were made on a commercial particle sizer for bigger particles (Fig S1): \SI{100}{\nano\meter}, \SI{500}{\nano\meter}, \SI{2}{\micro\meter} and \SI{5}{\micro\meter} diameter polystyrene particles and \SI{5.5}{\micro\meter} diameter PMMA particles.
Raw signals are represented together with the minimal SNR recommended by the manufacturer and the value SNR=1 (figure \ref{fig:S2}).
As expected for each size of particles, the scattering signal is proportional to the volume fraction (figure \ref{fig:S2}, left), simply because it goes linearly with the number of particle in the probe volume. 
However, for a fixed volume fraction, the signal is a non-monotonic function of the particle diameter.
This non-monotonic dependence is easily explained by the fact that, for a fixed volume fraction, the signal goes as the product of the number density of particle (which goes as the inverse of the particle volume) and the scattering cross-section (which increases as the squared volume in the Rayleigh regime and increases much less in the Mie regime).
When the particle diameter increases from 100 to \SI{500}{\nano\meter}, at the upper boundary of the Rayleigh regime, the signal practically increases by a factor $0.5\times10^4$, but a much smaller increase is observed for larger diameters (figure \ref{fig:S2}, right).
As explained by the manufacturer: for particles bigger than a micron in diameter, the number of particles inside the probing volume gets too low for a readable signal when volume fractions are less than $10^{-5}$.\\

Figure \ref{fig:S2} represents the results mentioned in the refractive index mismatch section. 
The size of particles was measured by CASS in both water and oil suspensions, and we show the corresponding intensity auto-correlation functions $|g^{(1)}(\tau)|^{2}$.
The CONTIN and CUMULANT fits used to retrieve the diffusion coefficients from the experimental auto-correlation function are also represented.

\setcounter{figure}{0}\global\def\thefigure{S\arabic{figure}}

\begin{figure*}\label{fig:S1}
\includegraphics[width=\linewidth]{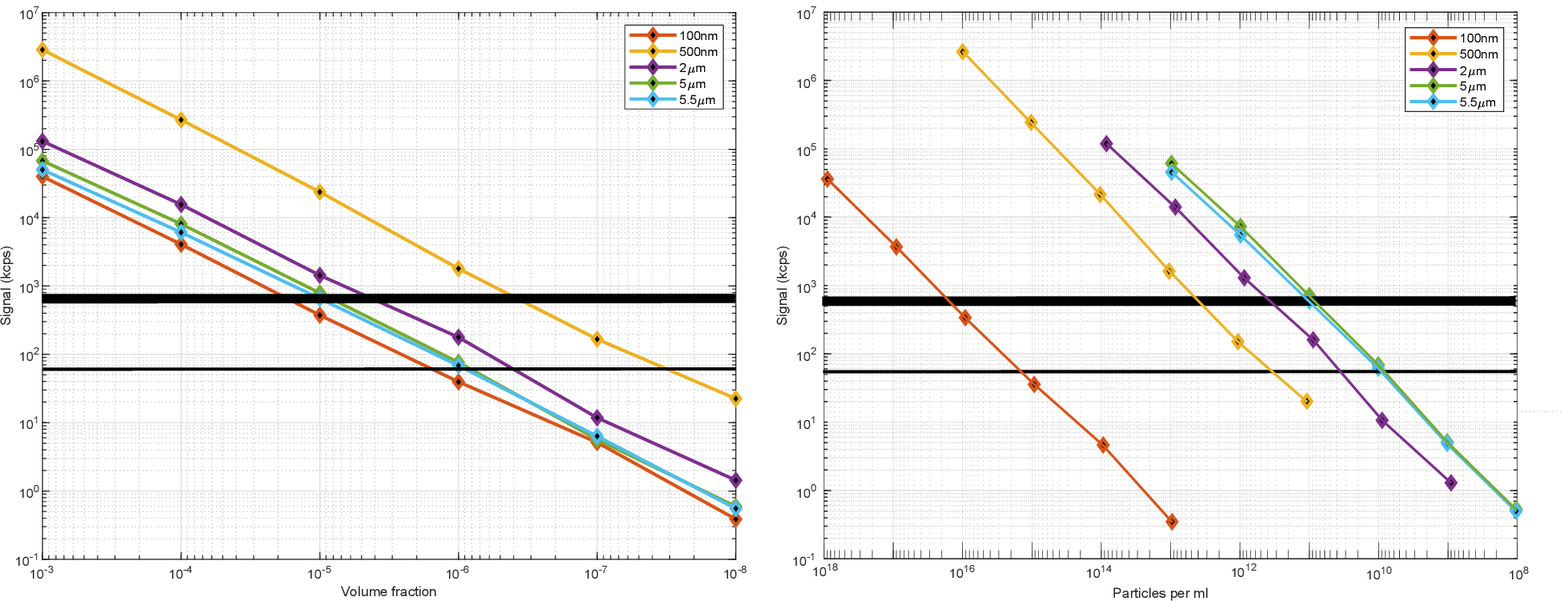}
\caption{\label{fig:6} (left) Signal (in kilo counts per second) from a commercial DLS particle sizer for polystyrene particles ($\SI{5}{\micro\meter}$, $\SI{2}{\micro\meter}$, $\SI{100}{\nano\meter}$, $\SI{500}{\nano\meter}$ diameters) and PMMA particles ($\SI{5.5}{\micro\meter}$ diameter) and volume fractions ranging from $10^{-8}$ to $10^{-3}$. 
The thick horizontal black line is the recommended minimum count (which gives an error of about $10\%$ on the estimated particle diameter) while the thin black line represents a signal to noise ratio of 1. 
(right) Same dataset represented as a function of the number of particles per mL.\\}
\end{figure*}

\begin{figure}
\includegraphics[width=\linewidth]{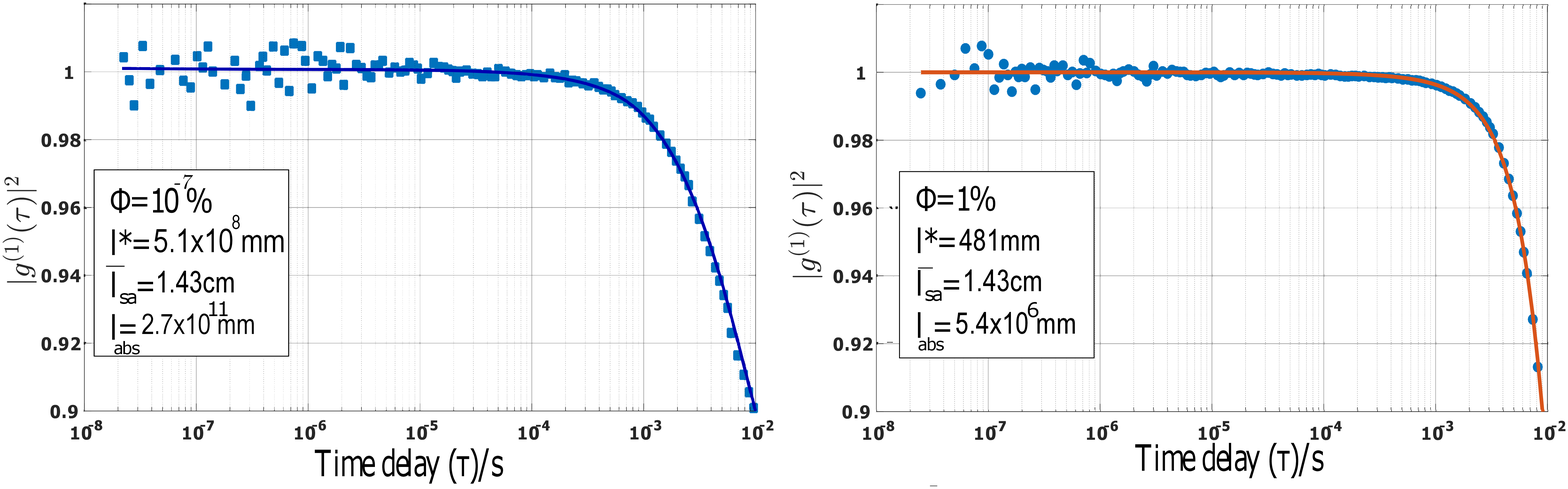}
\caption{\label{fig:S2} 
CASS auto-correlation function $|g^{(1)}(\tau)|^{2}$, shown (left) for a $10^{-9}$ volume fraction suspension of \SI{27}{\nano\meter} diameter polystyrene particles in water (black dots) and its CONTIN fit (blue line), and (right) for a $1\%$ volume fraction suspension of \SI{5.5}{\micro\meter} diameter PMMA particles (refractive index 1.488) in a 1.474 refractive index oil (blue dots) and its CUMULANT fit (red line)}
\end{figure}
\end{document}